# Defining causal mechanism in dual process theory and two types of feedback control

*Yoshiyuki Ohmura, Yasuo Kuniyoshi

The University of Tokyo

Abstract

Mental events are considered to supervene on physical events. A supervenient event does not change without a corresponding change in the underlying subvenient physical events. Since wholes and their parts exhibit the same supervenience–subvenience relations, inter-level causation has been expected to serve as a model for mental causation. We proposed an inter-level causation mechanism to construct a model of consciousness and an agent's self-determination. However, a significant gap exists between this mechanism and cognitive functions. Here, we demonstrate how to integrate the inter-level causation mechanism with the widely known dual-process theories.

We assume that the supervenience level is composed of multiple supervenient functions (i.e., neural networks), and we argue that inter-level causation can be achieved by controlling the feedback error defined through changing algebraic expressions combining these functions. Using inter-level causation allows for a dual laws model in which each level possesses its own distinct dynamics. In this framework, the feedback error is determined independently by two processes: (1) the selection of equations combining supervenient functions, and (2) the negative feedback error reduction to satisfy the equations through adjustments of neurons and synapses.

We interpret these two independent feedback controls as Type 1 and Type 2 processes in the dual process theories. As a result, theories of consciousness, agency, and dual process theory are unified into a single framework, and the characteristic features of Type 1 and Type 2 processes are naturally derived.

# 1. Introduction

Human cognition can exhibit two different modes of operation for the same stimulus. One is a fast, automatic, associative response, and the other is a slower, deliberative, rule-based reasoning (Sloman, 1996; Smith & DeCoster, 2000; Evans, 2008; Kahneman, 2011). Dual-process theory labels these two modes "Type 1" and "Type 2," widely known as "System 1" and "System 2." However, the existence of two distinct systems has been questioned, and recently, it has been suggested that the term "system" should be abolished and replaced with clusters of characteristics (C1/C2) or processing modes (Bellini-Leite, 2018).

At the behavioral and computational levels, the default interventionist theory (Evans & Stanovich, 2013)—where Type 1 provides default intuitive output and Type 2 monitors and logically corrects it—is empirically fruitful. However, evidence that people can intuitively perceive logical contradictions (De Neys, 2012; De Neys and Pennycook, 2019) supports a parallel/competitive dynamics between these two processes (De Neys, 2021; De Neys, 2023), rather than a serial handover. This is because perceiving logical inconsistency requires detecting a conflict between the outputs of the two processes. Discussions continue regarding the nature of these two types of processing, their integration, and their respective characteristics (Samuels, 2009; Bellini-Leite, 2022; De Neys, 2023).

Findings from cognitive neuroscience have not yet revealed distinct activity patterns reflecting differences between the two types of thinking. A recent meta-analysis suggests that the prefrontal control network (including the medial/superior prefrontal cortex, anterior cingulate cortex, insular cortex, and left inferior frontal gyrus) is involved during task performance based on dual process theory; however, the distinction between the two thinking modes remains ambiguous (Gronchi et al., 2024). This lack of clarity may support a hierarchical model where Type 1 and Type 2 processes share and utilize the same neural circuitry, rather than suggesting distinct neural bases for the two processes.

Such theoretical developments suggest that treating the Type 1/Type 2 distinction as a difference in computational methods rather than as fixed modules represents a promising direction (Solman, 1996; Bellini-Leite, 2022).

This paper proposes a Dual-Laws Model to explain dual-process phenomena. Dual-Laws Model was originally constructed as an inter-level causal model for theories of consciousness and agency (Ohmura & Kuniyoshi, 2025; Ohmura et al., 2026), but a similar hierarchical model has also been proposed for dual-process theory (Carruthers, 2019; Frankish, 2019). Our model focuses on causal mechanisms, leaving a gap with cognitive processes. Meanwhile, traditional dual-process theory is a cognitive theory, presenting the challenge that causal mechanisms and the interaction mechanisms between Type 1 and Type 2 processes remain unclear. Integrating both approaches is expected to compensate for each other's shortcomings.

The key idea is that, above the neural substrate, there exists a coarse-grained level governed by higher-level laws that independent of lower neural-level laws—*dual laws*—that (a) are independent of specific neural implementations, (b) can nonetheless exert downward causal influence on neural dynamics, and (c) naturally accommodate shared circuitry across tasks. Within this framework, Type 1 and Type 2 processes are defined by differences in the control mechanism of feedback errors, which are independently determined by both higher-level states and lower-level neural states. Our model directly addresses the specification problem by supplying an ontologically modest but causally potent level of theory.

## 2. Dual-Laws Model

We have proposed the *dual-laws model* as a framework for constructing a testable theory of consciousness (Ohmura & Kuniyoshi, 2025). Subsequently, we suggested that this model can also explain important features of agency (Ohmura et al., 2026).
The distinctive aspect of this model is that it attempts to solve the problems of mental causation (Davidson, 1970; Searle, 1980; Mayer, 2018) and agent causation (Chisholm, 1982; O'Connor, 2009; Steward, 2012) within the context of a physical system by appealing to an inter-level causation mechanism.

Although Kim's exclusion argument is often used to deny inter-level causation (Kim, 1998), since Kim does not actually assume the existence of higher-level hierarchies composed of multiple elements, our concept of inter-level causation falls outside the scope of his argument.

A causal mechanism (Salmon, 1985) is defined by a cause and a causal transmission mechanism. A cause must not directly bring about an effect without a causal transmission mechanism; that is, one must be able to intervene on the cause independently of the effect.

Ohmura et al. (2026) refer to a cause at the whole-system level that can be intervened upon independently as a *supervenient cause*, and mathematically clarify why such intervention is possible. Such independent intervention at the supervenience level enables the dual-laws model. Thus, although originally formulated to model the mechanism of inter-level causation, its similarity to dual process theory has not been explored. In this section, as preparation for discussing their relationship, we first present an inter-level causation mechanism and its formulation, then relate it to a theory of agency, and finally discuss its relationship to a theory of consciousness.

## 2.1 Inter-level Causal Mechanism

Mental events supervene on physical events. For example, the sensation of pain cannot occur without physical changes. In general, *supervenience* is defined as a relation in which there can be no change in the higher-level state without a change in the lower-base state. Since the whole supervene on its parts, some have argued that mental causation can be explained by downward causation from whole to parts (Sperry, 1991; Steward, 2012; Mayr, 2018).

A causal mechanism consists of a cause and a causal transmission mechanism. A cause cannot influence its effect directly without such a mechanism. Because the whole cannot change independently of its parts, changes in the whole cannot themselves serve as causes.

To address this, Ohmura et al. (2026) assume multiple supervenience–lower-base relations, and consider a causal mechanism from a supervenience level composed of multiple supervenient entities to the corresponding lower base level.

A key feature of our model is that the coarse-graining that constitutes the supervenience level is generalized from physical states to *functions*. By assuming that the supervenience level is composed of multiple functions (i.e., neural networks), we can define equations at this level through combinations of functions. In the case of the brain, supervenient functions can be interpreted as neural networks composed of neurons. The inter-level feedback control that modifies the lower-base level so as to satisfy the equations defined at the supervenience level corresponds to the causal transmission mechanism.

Equations at the supervenience level can be specified by index sequences if we assume that an index uniquely determines a supervenient function. Thus, changing the index sequence changes the equation, which in turn changes the feedback error. Meanwhile, because feedback error is computed using supervenient functions, it is also influenced by changes in the lower base level. In other words, feedback error can independently receive influences from both the supervenience level and the lower base level.

Changing the index sequence alters the equations that define the error, but changes of neural states at lower-base level do not occur without a causal transmission mechanism. The causal transmission mechanism involves inter-level feedback control to satisfy the equation through negative feedback control of neural states. Thus, changes in index sequences satisfy the conditions for a cause in the causal mechanism. From this, the supervenient cause and the causal transmission mechanism become clearly defined.

## 2.2 Formalization

We define base level states as $\mathrm{SUB}_i \subset \mathbb{R}^{n_i}$ for an index set $i \in I, I \subset \mathbb{N}$. We define the corresponding supervenient function whose domain is direct product of $N$ sets $\mathbb{R}^m \times \mathbb{R}^m \times \cdots \times \mathbb{R}^m$ and codomain is $\mathbb{R}^m$ as follows: $\mathrm{SUP}[N]\colon \mathbb{R}^m \times \mathbb{R}^m \times \cdots \times \mathbb{R}^m \to \mathbb{R}^m$.

We define the bridge function $\mathrm{B}\colon \mathrm{SUB} \to \mathrm{SUP}[N]$. Each supervenience-lower base relation can be represented by $X^i = b_i(x^i)$, $X^i \in \mathrm{SUP}[N_i], x^i \in \mathrm{SUB}, b_i \in \mathrm{B}, i \in I.$

Consider an index sequence $c = [i_0, i_1, \ldots] \in C$, where $i_k \in I$. Each index corresponds one-to-one with an element of SUP. Let error function $E\colon \mathbb{R}^m \to \mathbb{R}^m$ and $S\colon C \to E$ be a mapping from sequence to error function.

Array of index sequence $\boldsymbol{c} = [c_0, c_1, \ldots]$ is a supervenience level state. The change in the state is supervenient cause.

Let array of error functions: $\boldsymbol{er} = [er_0, er_1, \ldots] \coloneqq [s(c_0), s(c_1), \ldots] = \varepsilon(\boldsymbol{c})$, where $er_k \in E, c_k \in C, s \in S$. Here, $\coloneqq$ is Pearl's assignment operator, which means left side "is determined by " right side (Pearl, 2009). We can use the assignment operator to show the causal asymmetric relationship because the index sequences change without a change in error functions. A unique feature of our formulation is the use of the semantics of causality. If we do not introduce the asymmetry between cause and effect, then our formulation cannot be distinguished from a standard single-level dynamical system model in the semantics of physics. We should distinguish the semantics of physics and the semantics of causality.

The error function is independently influenced by both changes in lower base and changes in the index sequence at the supervenience level. We define feedback error $\varepsilon(\boldsymbol{c})(d), d \in \mathbb{R}^m$. In inter-level feedback control, the lower base states are controlled to satisfy an equation: $\varepsilon(\boldsymbol{c})(d) = 0$.

## 2.3 Characterizing Agency

In the supervenient causation mechanism, the supervenience level changes index sequences, while the lower base level changes physical states such as neural activity and synapses. A model that assumes different dynamical laws for the supervenience level and the lower base level is referred to as the *Dual-Laws Model*.

If we call the dynamical laws of the lower base level Dynamics 1, then neural states are influenced by feedback error (itself constrained by the supervenience level) and by bodily or environmental inputs. If we call the dynamical laws of the supervenience level Dynamics 2, then these laws modify and select the index sequences. The neural mechanism determining Dynamics 2 plays the role of the "I." Through this mechanism, changes in the index sequences generate changes in feedback error, producing downstream changes in the behavior of the lower base (Figure 1). This top-down causal transmission explains the initiation of an act (Ohmura et al., 2026).

This model satisfies the three core conditions for agency proposed by Barandiaran et al. (2009):

- Individuality: The model avoids infinite regress by positing an agent-specific supervenience-level dynamics. The characteristics of an agency are determined independently of external observers.
- Asymmetric relation: Because supervenience-level interventions temporally precede and independently influence the lower base, the relation is causal and asymmetric.
- Normative: Lower-base control is implemented via goal-directed feedback regulation of neural states, thus fulfilling the normative criteria.

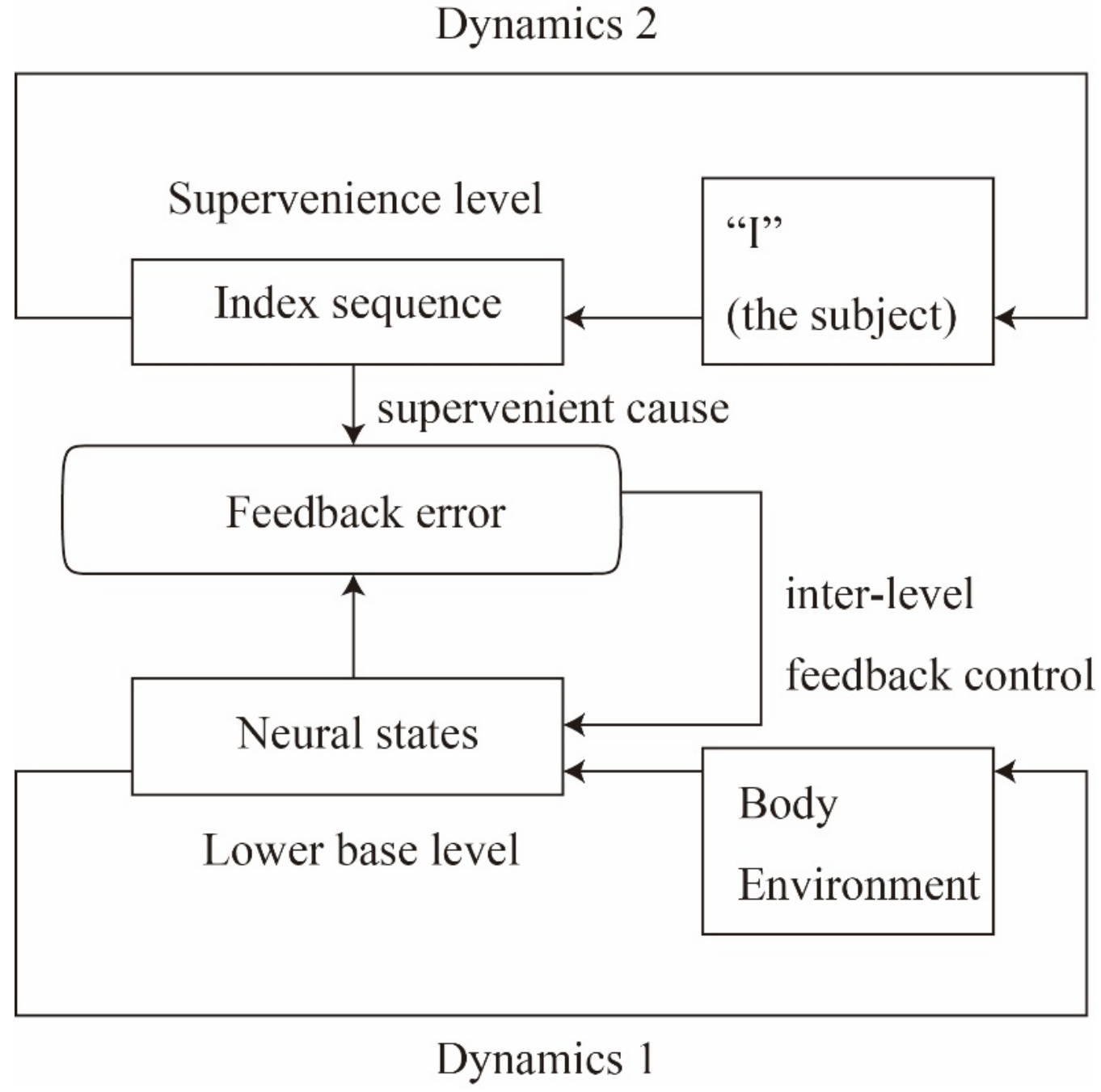

Figure. 1 Dual-Laws Model. The supervenience level is composed of multiple supervenient functions. The feedback error is the error involved in satisfying the equations defined by combining these supervenient functions (i.e., neural networks). These equations vary according to the index sequences. The dynamical laws at the supervenience level modify and select these index sequences, and these dynamical laws play the role of the "I." At the same time, the feedback error can be adjusted at the lower base level by neurons, synapses, and other components that constitute the supervenient functions. In this way, the feedback error is independently influenced by the index sequences at the supervenience level and by the neuronal states at the lower base level, which are open to environmental input, making the dual-laws model possible.

## 2.4 Characterizing Conscious Processing

Ohmura & Kuniyoshi (2025) argue that a testable theory of consciousness must acknowledge the causal efficacy of consciousness to characterize the functional aspects of consciousness. Here, rather than focusing on the causal efficacy of qualia, we have attempted to model the mechanism of consciousness generation and the causal efficacy of the resulting consciousness within a physical system characterized by inter-level causality. Since consciousness typically supervenes on physical processes, attention is directed to downward causation from whole to parts (Sperry, 1991). However, the causal efficacy of the supervenience level over the lower base is not identical to the causal efficacy of consciousness itself.

We propose that supervenient causation gives rise to neural activity correlated with conscious experience. In order to explain causal efficacy of consciousness, we assume that this neural activity influences the mechanism corresponding to the "I," which in turn influences Dynamics 2, thereby allowing conscious content to indirectly influence the lower base level through supervenient causation after generating the consciousness.

We suggest that Dynamics 2—the mechanism for modifying index sequences—is carried out by neural mechanisms corresponding to the "I." Importantly, the "I" serves both as the initiator of an act and as the experiencer. Although this unification is intuitive, models without inter-level causation mechanism struggle to explicitly include such an "I."

Our model concerns causal mechanisms, not specific cognitive or psychological architectures. Dual process theories invoking hierarchical structures (Carruthers, 2009; Frankish, 2009) make the relationship with the Dual-Laws Model particularly meaningful. Meanwhile, dual process theory, originally proposed to interpret findings from psychological experiments, lacks clear causal mechanisms. Therefore, integrating the Dual-Laws Model with dual process theory is expected to yield substantial synergy. The aim of this work is to bridge the gap between them and develop an integrated theory.

# 3. Bridging the Gap Between the Dual-Laws Model and Dual Process Theory

In this section, we propose a hypothesis explaining dual processes using the framework of the Dual-Laws Model. Dual process theory posits the existence of two distinct processes: one that is intuitive and fast, and another that is rule-based and slow. Here, we attempt to explain dual processes by considering two different methods for reducing feedback error. One is a method that continuously reduces feedback error through feedback control of neural activity at the lower base level (Type 1). The other is a method that discretely reduces feedback error by modifying or selecting the equations that define the feedback error at the supervenience level (Type 2). In order to consider Type 2 adjustments, the relationships among multiple equations must first be defined.

## 3.1 Competitive Relationship

We define the supervenient cause as the process that changes the index sequences used to define equations at the supervenience level. If an equation is uniquely determined by an index sequence, then the index sequence and the equation can be treated interchangeably. Under this assumption, how are equations that define feedback error selected at the supervenience level? It is not necessarily the case that only one equation is selected at any given time; rather, we assume that multiple equations can be selected simultaneously.

At this point, the evidence suggests that, according to dual-process theory, different responses to the same stimulus require a group of competing equations. We assume that the equations selected at the supervenience level determine responses. When mutually exclusive equations cannot be satisfied at the same time, competitive relations can be defined. Thus, we define exclusivity between equations.

A simple method for introducing such mutually exclusive relations is to impose a group structure on the set of supervenient functions. A group is an algebraic structure defined by a set and a binary operation, along with an identity element and an inverse element. The identity and inverse elements can be used to define counterfactuals. For example, if an object moves to the right, then it is not moving to the left; movement to the right and left are inverses. By introducing a group structure, mutually exclusive relations can be defined and non-independent equations can be constructed.

Furthermore, group structures are generally defined by axioms, which themselves are equations. Since these equations can also be defined using index sequences, our model lends itself naturally competitive relationships.

## 3.2 Equation Network

In the previous section, we defined competitive relations among equations. However, competition alone merely suppresses equations that are mutually exclusive, and it becomes difficult for conflicts—of the sort described in dual process theory—to emerge. Since the competitive relations are clearly defined by identity and inverse elements, one could simply select equations that are not in competition from the start. For this reason, we consider associative relationships among equations. Through association or prediction, the likelihood of certain sets of equations being selected together can be increased. If equation selection is determined by at least two distinct mechanisms, then it becomes possible for equations that should be mutually exclusive to be selected simultaneously.

Although this study does not address how such an equation network is constructed, we assume the existence of a mechanism that reduces feedback error through adjustments to this equation network. With this, it becomes possible to introduce two types of error-reduction methods: one that adjusts feedback error from the supervenience level and another that reduces feedback error from the lower base level.

### 3.3 Two Types of Feedback Control

We consider the two methods of feedback-error reduction as corresponding to Type 1 and Type 2 in dual process theory.

**Type 1** is *inter-level feedback control*, a mechanism that adjusts base-level neural activity to satisfy the equations defined at the supervenience level. This process is relatively automatic, but it is not a feedforward computation but goal-directed activity that satisfies constraints from the supervenience level. It is a process that modifies input-output relations of the supervenient functions and is assumed to allow continuous control.

**Type 2** is a process that reduces feedback error by changing the selection of a set of equations at the supervenience level. The equations are adjusted by modifying the index sequence. This process is inherently discrete and is assumed to involve exploratory or algorithmic adjustments. The qualitative difference between Type 1 and Type 2 lies in the fact that the former continuously adjusts the functions themselves, while the latter requires discrete control through reordering of index sequences or selecting a group of index sequences.

These two methods of control can be naturally incorporated when assuming the Dual-Laws Model, and there is no reason to believe that Type 2 exists only in humans. However, since Type 2 is inherently discrete, it is plausible that it becomes particularly advanced in humans, who have developed sophisticated linguistic capacities.

## 4. Two Types of Belief Networks

In the previous section, we assumed a network of equations at the supervenience level. Each equation is composed of shared supervenient functions, and associative as well as exclusive relations are defined among these equations.

## 4.1 Embodied Beliefs

Here, if we focus on the fact that equations define equivalence relations, then these equations can be regarded as *beliefs*. If these equations induce bodily actions that satisfy them, then these beliefs can also be viewed as *desires*. We therefore refer to the equations embedded within the network at the supervenience level, introduced in the previous section, as *embodied beliefs*.

Embodied beliefs are assumed to be formed based on perceptual experience. Although these beliefs are discrete and symbolic, they must be distinguished from language-based beliefs. Embodied beliefs are directly tied to subjective perceptual experience and are bodily embedded; even animals without language can possess them. Furthermore, since these beliefs are functions used in automatic computation of feedback error, the computation itself is not related to consciousness.

Assuming such embodied beliefs as pre-linguistic beliefs is extremely important for understanding the evolution of language. If language is considered something that is passively provided by society, an infinite regress arises regarding where the first language originates. Thus, it is natural to assume that organisms already possessed a symbolic capacity prior to language.

## 4.2 Linguistic Beliefs

If embodied beliefs can induce bodily actions, they can be viewed as desires. Here, we assume that embodied beliefs become connected to linguistic expressions such as gestures, spoken language, or written symbols. These externally expressed symbolic sequences, when processed through sensory feedback, allow the corresponding perceptual experience to be consciously reflected upon (Jackendoff, 1996; Carruthers, 2002; Jackendoff, 2007; Prinz, 2023).

Our hypothesis is that sensory information becomes available to consciousness through structuring by Type 1 processes (Ohmura & Kuniyoshi, 2025). Based on this hypothesis, it should be possible to consciously access self-generated symbolic sequences. We consider that the conscious awareness of linguistic expressions corresponding to pre-linguistic embodied beliefs serves an important function.

Evidence for higher-order cognition occurring unconsciously is scarce (Pournaghdail & Schwartz, 2020), suggesting that conscious processing is necessary for semantic integration (Moors et al., 2017), semantic association (Skora, et al., 2021) and cognitive control (Ansorge et al., 2011; Huang & Li, 2025; Pedale et al., 2025). These findings suggest that performing Type 2 processes completely unconsciously may be difficult.

If unconscious associations and conditioning are restricted, the construction of predictive models should rely heavily on Type 2 processes, rather than Type 1 processes. For example, the reduction in cognitive effort accompanying mastery of motor skills supports this. That is, predictive models should possess not only statistical properties but also discrete, meaningful structures at the supervenience level.

At this point, we can distinguish two types of embodied beliefs:
(1) *classical embodied belief*: embodied beliefs directly tied to subjective perceptual experience and responsible for generating symbolic expressions such as language or gesture, and
(2) *linguistic beliefs*: embodied beliefs responsible for consciously perceiving the sensory feedback of generated symbolic expressions.

These linguistic beliefs are influenced by the linguistic system formed through social interaction. Thus, their structure differs from that of classical embodied beliefs formed individually through perceptual experience. Human uniqueness arises from the fact that, in addition to the classical embodied belief system, a linguistic belief system influenced by language is added.

Stanovich & Toplak (2012) proposed *cognitive decoupling* as a feature of Type 2 in dual process theory. They referred to representations directly influenced by perceptual experience as *primary representations*, and those influenced by language as *secondary representations*, linking the simulation function using the latter to Type 2. In our framework, the classical embodied beliefs and the linguistically influenced linguistic beliefs correspond to primary and secondary representations, respectively.

In the next section, we characterize the two processes based on the hypotheses developed thus far.

# 5. Characterizing Type 1 and Type 2 Processes

Up to this point, we have introduced concepts necessary to align the Dual-Laws Model with dual process theory. The Type 1 process is a mechanism that reduces feedback error—defined at the supervenience level—through control at the lower base level. The Type 2 process reduces feedback error by modifying and selecting the equations that define the feedback error. At the same time, we introduced *embodied beliefs* as the basis for setting equations at the supervenience level. Humans are assumed to possess an extended subset of embodied beliefs—*linguistic beliefs*—that reflect linguistic structure. While this belief network constitutes a single system, the correspondence between perceptually influenced classical embodied beliefs and their linguistically influenced counterparts creates the possibility of human-specific interactions. Traditional dual process theory can be interpreted as focusing on phenomena arising from contradictions between classical embodied beliefs and linguistic beliefs.

Below, we examine features discussed in past dual process theories and evaluate whether they serve as defining features or merely correlations, from the standpoint of our framework.

*Fast and Slow*

In our model, the Type 1 process must operate faster than the Type 2 process. Because Type 1 feedback control is involved in goal-directed behavior, if the Type 2 process were to change goals before Type 1 converges, consistent goal-directed behavior would be impossible. Therefore, Type 1 processes must fundamentally be faster, whereas Type 2 processes must be slower.

*Predictive and Symbolic*

Bellini-Leite (2022) considered Type 1 to be statistical and predictive, and Type 2 to be symbolic. Although prediction is necessary for feedback control in Type 1 processes, we do not regard predictiveness as a defining feature of Type 1. In contrast, Type 2 involves reducing feedback error by adjusting discrete symbolic equations, making it reasonable to characterize Type 2 as symbolic.

As mentioned earlier, based on the finding that associative learning in the unconscious is restricted (Pournaghdail & Schwartz, 2020), we consider that Type 2 processes are involved in the learning of predictive models.

*Intuitive and Rule-Based*

Type 1 is often described as intuitive (Evans, 2017), and this can be interpreted as judgments strongly influenced by prelinguistic classical embodied beliefs. Indeed, if the Type 2 process is inactive and feedback error is reduced solely by the Type 1 process, reflective processes such as inner speech (Vicente & Manrique, 2011) do not arise, making decisions appear intuitive. However, because Type 1 processes are influenced by the belief network, logical or rule-based judgments are possible (De Neys, 2019). These characteristics should be seen as tendencies rather than strict distinctions.

*Automatic and Controlled*

Stanovich & Toplak (2012) proposed automaticity as a defining feature of Type 1. Type 1 process is indeed relatively automatic. However, we distinguish it from feedforward computation processes. Type 1 is feedback control that aims to satisfy goals determined at the supervenience level. Crucially, Type 1 should *not* be considered uncontrolled. Melnikoff and Bargh (2018) criticized the view that Type 1 processes are uncontrollable.

*Unconscious and Conscious*

Type 1 processes are necessary for generating consciousness, but we consider that these processes themselves do not require consciousness. On the other hand, if it is correct that the semantic integration and association capability of unconscious processing is restricted, then Type 2 processes are predicted to fundamentally require consciousness.

*Personal Experience vs. Culture*

Stanovich & Toplak (2012) emphasized cognitive decoupling as a feature of Type 2. We also consider classical embodied beliefs to be strongly tied to personal perceptual experience, whereas linguistic beliefs are strongly tied to culture. Human linguistic systems exert a strong influence on Type 2 processes. In particular, competition between classical embodied beliefs and linguistic beliefs may generate situations where Type 1 cannot reduce feedback error, requiring control at the supervenience level via Type 2. This is because Type 1 fundamentally cannot resolve certain conflicting states. We consider that Type 2 processes typically require cognitive decoupling.

*Parallel and Sequential*

Both Type 1 and Type 2 processes share the neural network at the lower base level. There is no inherent reason why Type 2 must be sequential. However, human language systems are highly sequential. The processes of expressing embodied beliefs as symbolic sequences and making these sequences conscious—through inner speech, for example—introduce sequentiality. Thus, Type 2 tends to appear sequential as a result.

*Evolutionary Old and Evolutionary Recent*

Both Type 1 and Type 2 processes may exist in non-human animals. However, in humans, the evolution of language likely allowed Type 2 processes to develop in a highly specialized and advanced manner.

*Efficient vs. Inefficient*

Type 1 can produce continuous changes, whereas Type 2 involves discrete, often exploratory adjustments; thus, Type 1 is generally more efficient. However, efficiency is highly task-dependent and therefore should not be considered a defining feature.

On the other hand, we believe that Type 1 processes are necessary for the generation of consciousness, and that Type 1 processes occur simultaneously with Type 2 processes. Therefore, the discussion of consciousness and efficiency in dual-process theory presents a complex picture (Melnikoff & Bargh, 2018) and is difficult to simplify.

*Subpersonal Level and Personal Level*

Frankish (2019) and Carruthers (2019) regarded Type 1 and Type 2 as operating at different hierarchical levels, which aligns with our model. However, they do not characterize Type 1 and Type 2 as different types of computation mechanisms to reduce a common feedback error. Moreover, they tend to view Type 1 as an evolutionarily old and fixed neural circuit. In our model, Type 1 neural circuits are adjusted according to feedback error and are therefore not fixed. Because the supervenience and lower base levels share the same physical substrate, highly advanced processing would be impossible if Type 1 circuits were fixed.

## Summary of Defining Features

We regard Type 1 as fast inter-level feedback control that includes goal-directed behavior. Type 2 is symbolic and relatively slow, and in humans, highly developed due to linguistic beliefs. Cognitive decoupling is a typical Type 2 process.

Crucially, Type 1 and Type 2 processes are different types of computational mechanism to reduce a shared feedback error**.** The most direct method for reducing error is the Type 1 process, which must therefore be relatively fast to support goal-directed behavior. Type 2 involves algorithmic control over discrete states.

Finally, Type 1 and Type 2 are not mutually exclusive. Human behavior arises through their interaction, and experimental results should not be assumed a priori to reflect only one or the other. Furthermore, because the supervenience and lower base levels share the same physical substrate, regional neural activation does not necessarily map cleanly onto Type 1 vs. Type 2 distinctions. Although the relationship between the Dual-Laws Model and the actual experimental results in the dual process theory is not simple, our model nonetheless accounts for two types of cluster effects from the integration of different kind of feedback control processes.

# 6. Conclusion

Our approach differs from conventional methods that construct theories from experimental results. We clarified the conditions under which *supervenient causation* becomes possible—namely, conditions satisfying (1) that mental events supervene on physical events, meaning that mental events do not change without changes in physical events, and (2) that mental events possess causal efficacy over physical events (Ohmura et al., 2026). Causal relations must involve both a cause and a causal transmission mechanism. By defining hierarchical levels through multiple supervenience–lower-base relations, we identified the causes that arise at the supervenience level and the causal transmission mechanisms through which they affect the lower base. As a result, we proposed a Dual-Laws Model in which distinct dynamics exist at the supervenience level and the lower base. The Dual-Laws Model is a physical system model with inter-level causation mechanisms to explain functional aspects of mental processes.

This dual-laws model was originally developed to elucidate the conditions for the emergence of consciousness (Ohmura & Kuniyoshi, 2025) and the initiation processes of an agent's actions (Ohmura et al. 2026). However, its relationship to dual process theory has not been previously examined. Theories of consciousness and agency focus on causal mechanisms and are highly abstract, which has made their relationship to cognitive processes unclear. We therefore expected that discussing their relationship with dual process theory would create a synergistic effect. The Dual-Laws Model makes the connection with cognitive processes more explicit, while dual process theory gains mathematical grounding for characterizing Type 1 and Type 2 processes. If successful, this integration would allow us to construct a unified theory that brings together a theory of consciousness, a theory of agency, and the dual process theory.

In this study, we clarified the relationship between the Dual-Laws Model and dual process theory. As a result, Type 1 and Type 2 processes in dual process theory can both be understood as feedback control mechanisms, distinguished by the manner in which they reduce feedback error. Type 1 reduces feedback error by adjusting supervenient functions through changes in neuronal and synaptic states, while Type 2 reduces feedback error by modifying and selecting the multiple equations used to define the error itself. The conflict traditionally assumed to mediate switches between Type 1 and Type 2 in dual process theory corresponds to feedback error. Our model treats Type 1 and Type 2 as non-exclusive processes that both aim to reduce the same feedback error.

Because the Type 2 process defines the feedback error, the Type 1 process must operate more rapidly; otherwise, consistent goal-directedness cannot be achieved. At the same time, introducing exclusivity among multiple equations allows us to define mutually incompatible equations that cannot be satisfied simultaneously. This creates situations in which Type 1 control alone cannot reduce feedback error, making the Type 2 process necessary.

However, exclusivity introduced solely through group algebraic structures is insufficient: although it prevents simultaneous selection of competing equations, it also removes the need for the Type 2 process. Thus, exclusivity among equations must be supplemented with an associative process between them. With two distinct processes present, contradictions may arise in equation selection, and resolving these contradictions is the role of the Type 2 process.

In humans, advanced linguistic beliefs can be seen as a part of embodied beliefs formed through social experience. Because social structures and linguistic structures influence them, conflicts can arise between linguistic beliefs and classical embodied beliefs formed through personal perceptual experience. In such cases, the Type 2 process is required to resolve the conflict. This can be linked to cognitive decoupling and working memory (Stanovich & Toplak, 2012; Evans & Stanovich, 2013). While the development of language has increased the importance of the Type 2 process in humans, we do not consider the process itself to be uniquely human.

Behavioral and cognitive experiments are generally influenced by both Type 1 and Type 2 processes. Therefore, it may be difficult to isolate the pure influence of either process from experimental data. Moreover, in our hierarchical model, most Type 1 and Type 2 processes rely on shared neural circuits, making it difficult to distinguish between them using coarse measurements such as activated brain regions.

We originally proposed the Dual-Laws Model with the aim of clarifying the conditions under which mental phenomena supervening on physical phenomena possess causal efficacy, thereby addressing a theory of consciousness and the homunculus problem. Given that a hierarchical dual process theory had already been proposed (Carruthers, 2019; Frankish, 2019), we believed that integration was possible. The important point is that theories developed for entirely different purposes can nonetheless explain various aspects of dual process theory. For example, our model clarifies why Type 1 must be relatively fast. It also explains why Type 2 is discrete—this follows from the fact that the supervenience level must be composed of multiple supervenient functions in order to define supervenient causation. In contrast, the characteristics of Type 1 should not be interpreted as autonomous feedforward computation; rather, they should be understood as feedback control. The sequential nature of Type 2 arises from linguistic influence and is therefore not essential. Type 1 is not necessarily associative, nor are Type 1 and Type 2 mutually exclusive; exclusivity is instead a constraint in decision-making. Humans can make embodied beliefs—normally inaccessible to consciousness—accessible through language, and inner speech clearly reveals the connection between consciousness and the Type 2 process. We have proposed that the Type 1 process itself is necessary for generating the contents of consciousness (Ohmura & Kuniyoshi, 2025), and because Type 1 and Type 2 are not inherently exclusive, the relationship between dual process theory and consciousness is not straightforward.

By integrating these two distinct theories, we have clarified the differences between Type 1 and Type 2 processes and characterized the features of both processes. This integration offers valuable insights into both theories. A remaining challenge is to elucidate the mechanism underlying the formation of embodied belief networks, the mechanism of equation selection at the supervenience level, and integration with AI models (Gronchi & Perini, 2024; Bellinit-Leite, 2024).

## Acknowledgement

## Funding

This research was supported by the JSPS KAKENHI (25H00448), Japan. The funding sources had no role in the decision to publish or prepare the manuscript.

## Author Contributions

YO: conceptualization, formalization, draft writing, and revision; YK: supervision.

## Statement and Declarations

The authors declared no potential conflicts of interest with respect to the research, authorship, and/or publication of this article.